# BIOSIGNATURES FROM EARTH-LIKE PLANETS AROUND M DWARFS

(Astrobiology, in press)

ANTÍGONA SEGURA,[1,*,†] JAMES F. KASTING,[1,*] VICTORIA MEADOWS,[2,*] MARTIN COHEN,[3,*] JOHN SCALO,[4] DAVID CRISP,[5,*] REBECCA A.H. BUTLER,[5,*] and GIOVANA TINETTI[6,*]

[1]Department of Geosciences, The Pennsylvania State University, University Park, Pennsylvania.
[2]Infrared Processing and Analysis Center and [5]NASA Jet Propulsion Laboratory, California Institute of Technology; and [6]California Institute of Technology/National Research Council, Pasadena, California.
[3]Radio Astronomy Laboratory, University of California, Berkeley, Berkeley, California.
[4]Department of Astronomy, University of Texas, Austin, Texas.
[*]Members of the NASA Astrobiology Institute.
[†]Present address: Infrared Processing and Analysis Center, California Institute of Technology, Pasadena, California.

## ABSTRACT

Coupled one-dimensional photochemical-climate calculations have been performed for hypothetical Earth-like planets around M dwarfs. Visible/near-infrared and thermal-infrared synthetic spectra of these planets were generated to determine which biosignature gases might be observed by a future, space-based telescope. Our star sample included two observed active M dwarfs—AD Leo and GJ 643—and three quiescent model stars. The spectral distribution of these stars in the ultraviolet generates a different photochemistry on these planets. As a result, the biogenic gases $CH_4$, $N_2O$, and $CH_3Cl$ have substantially longer lifetimes and higher mixing ratios than on Earth, making them potentially observable by space-based telescopes. On the active M-star planets, an ozone layer similar to Earth's was developed that resulted in a spectroscopic signature comparable to the terrestrial one. The simultaneous detection of $O_2$ (or $O_3$) and a reduced gas in a planet's atmosphere has been suggested as strong evidence for life. Planets circling M stars may be good locations to search for such evidence.

# INTRODUCTION

L OW-MASS MAIN SEQUENCE STARS of spectral class M (M dwarfs, or dM stars) are the most abundant stars in the galaxy, representing about 75% of the total stellar population. Relative to our Sun, M stars have many unusual characteristics. Their low masses (0.08–0.8 $M_\odot$) allow them to have main sequence lifetimes on the order of $10^{11}$ years (or more), *i.e.*, significantly longer than the current age of the universe. Because of their low effective temperatures their spectra are dominated by molecular absorption bands that redistribute their radiated energy in a distinctly non-blackbody fashion. Most M stars also exhibit significant temporal variability as a consequence of phenomena occurring within the region from their photospheres to their coronae. As a result, they emit large amounts of short-wavelength (SW) ultraviolet (UV) radiation and x-rays during their active periods.

Given these stellar characteristics, and the likely general characteristics of planets within the habitable zone (HZ) of an M star, the potential habitability of planets around M dwarfs has been questioned by a number of authors (*e.g.*, Dole, 1964; Kasting *et al.*, 1993). We discuss some of the apparent problems below. It seems likely that planets do form around such stars, and there is new observational evidence to support this. Numerical simulations predicted that Neptune-like and terrestrial-type planets may be common around M stars, given sufficient disk mass (Wetherill, 1996; Laughlin *et al.*, 2004), though formation of gas giants may be inhibited (Laughlin *et al.*, 2004). This prediction has been borne out by the detection of a Neptune-size planet (~18 $M_\oplus$) around the M2.5 star GJ 436 (Butler *et al.*, 2004). Because of the small mass of M stars, Neptune-like and possibly even terrestrial planets can be detected using the radial velocity and transit photometry techniques from ground-based telescopes (Doyle and Deeg, 2002; Kürster *et al.*, 2003; Hatzes, 2004; Sozzetti *et al.*, 2004). Recently, an even smaller (~7.5 $M_\oplus$), possibly rocky, planet has been discovered around the dM4 nearby star, GJ 876 (E.J. Rivera *et al.*, manuscript submitted for publication). Presently, three projects are searching for planets around M stars: the radial velocity program using the Hobby-Eberly telescope (Endl *et al.*, 2003); the astrometric program Stellar Planet Survey (STEPS) (Pravdo *et al.*, 2004, 2005), which uses the Palomar 5-m telescope; and the Keck radial velocity program (Marcy *et al.*, 1998).

Although Earth-mass planets around M stars—or any main sequence star besides the Sun—have yet to be discovered, we assume for now that these planets will eventually be found around M stars.

The HZ, as defined by Hart (1978) and Kasting *et al.* (1993), is the region around a star where a planet can maintain liquid water at its surface. For an M dwarf, the HZ is so close to the star that the planets are likely to become tidally locked within a relatively short time after they form (Dole, 1964; Kasting *et al.*, 1993). A planet with this characteristic, termed a "synchronous rotator," receives starlight on only one hemisphere, while the other hemisphere remains forever in the dark. If the dark side of the planet gets too cold, then the volatile compounds that make up the atmosphere and oceans may freeze out to form a giant ice cap. This process is sometimes termed "atmospheric collapse." Detailed models have shown that this is not an insurmountable impediment to planetary habitability for planets with even modestly dense atmospheres (Haberle *et al.*, 1996; Joshi *et al.*, 1997; Joshi, 2003). Given an insolation equal to that on Earth, a 100-mbar $CO_2$ atmosphere (~1/10[th] of the Earth's atmospheric mass) would transport sufficient heat to the dark side to avoid atmospheric collapse. Earth-like planets that are relatively far out in the HZ of M stars are likely to have high $CO_2$ concentrations as a consequence of the carbonate–silicate cycle feedback (Kasting *et al.*, 1993). High $CO_2$ levels increase the radiative cooling time of the atmosphere, thereby allowing heat to be transferred effectively from one hemisphere to another. If an extensive ocean is present, even a low-$CO_2$ planet like Earth ($f_{CO2} = 300$ ppmv) might remain clement, provided that a vigorous atmospheric hydrological cycle transports heat from the dayside to the nightside (Joshi, 2003).

In our Solar System, Venus and Titan are examples of nearly tidally locked bodies with dense super-rotating atmospheres. The surface temperature on Venus is believed to be nearly constant on both the day- and nightside. For Titan, equator-to-pole gradients in the troposphere could be as large as 20°, according to the information available at present (not including the most recent Cassini-Huygens results). However, the atmospheres on both these planets have significantly normalized the surface temperatures, even though Titan is a synchronous rotator with Saturn, and Venus rotates with a 243-day period (yet the atmosphere rotates once every 4 days). It is important to note that not all close-in planets become synchronous rotators. Mercury, for example, is in a 3:2 spin/orbit resonance, meaning that it rotates three times for every two revolutions around the Sun.

Another question with regard to the potential habitability of M-star planets is a deficiency in photosynthetically active radiation (PAR), which lies predominantly in the wavelength range 400–720 nm. Photosynthesis can proceed at wavelengths longer than 720 nm, but it does not produce oxygen (Heath *et al*., 1999). PAR emitted by M stars is diminished by the large number of photospheric absorption bands, notably those of molecular TiO between 459 nm and 625 nm (Heath *et al*., 1999, Fig. 4), and the redder bands near 690 and 705 nm. But this problem may be partially compensated by the perpetual illumination that a photosynthetic organism would receive on the dayside of the tidally locked planet (Heath *et al*., 1999). Furthermore, the minimum PAR flux required to sustain marine photosynthesis is only $\sim 5 \times 10^{-4}$ of the average flux incident at the Earth's surface (McKay, 2000). By comparison, the typical PAR flux for a habitable planet circling one of the M stars studied here is about 0.1 times Earth's flux. Hence, the lack of PAR should not prevent M-star planets from harboring $O_2$-producing organisms.

A potentially more serious problem is the temporal variability of M dwarfs. The most prominent features of this activity are flares—sudden, unpredictable releases of energy that range from "microflares," with energies as small as can be detected ($\sim 10^{28}$ erg of radiative energy release), to flares with blue and UV energies as large as $10^{34}$–$10^{37}$ ergs in the most active dMe stars [Hα emission line main sequence M stars (see Hawley and Pettersen, 1991; Liebert *et al*., 1999; Güdel *et al*., 2003, Table 1)]. Comparably energetic x-ray flares have been observed in other M stars [*e.g*., AU Mic (Cully *et al*., 1993), EV Lac (Favata *et al*., 2000)]. For the strongest flare stars, it is estimated that flares more energetic than $\sim 10^{32}$ ergs (the largest energies observed so far for solar flares) occur at a rate of roughly one per day, depending on their overall coronal luminosity (Audard *et al*., 2000, Fig. 4). Each flare is a unique event that can evolve on time scales from tens of seconds to a day or more (Houdebine, 2003). The frequency and lifetime of the flares depend on the energy of each event: large flares occur less often but last longer than the small ones (Gershberg, 1989).

In an Earth-like atmosphere, the energy released by x-rays can be absorbed and re-emitted, cascading down to longer wavelengths that affect ozone ($O_3$) chemistry in the stratosphere and below (Smith *et al*., 2004a,b). M dwarfs are generally much more active than solar-like stars (Gershberg *et al*., 1999), and so it is not easy to extrapolate the consequences of that activity to biological evolution on an M-star planet. For illustration of the more intense activity of M stars relative to solar-like stars of the same age, we compare the ROSAT soft x-ray coronal luminosities of α Cen (G2V+K1V) and its distant binary companion Proxima Cen (M5.5Ve). According to Huensch *et al*. (1998) $L_x \sim 10^{20}$ W for α Cen, while $L_x \sim 10^{21}$ W for Proxima Cen (Güdel *et al.*, 2004, Table 4). Considering that Proxima Cen is at least a factor of $10^{-2}$ fainter than α Cen in bolometric luminosity, this shows that $(L_x/L_{bol})$ is at least $10^3$ times larger for Proxima Cen. The M star Proxima Cen would be absolutely and effectively (for a planet in the liquid water HZ) a stronger source of x-rays than the G star α Cen, which has the same age.

Yet another problem for M-star planets is the possible erosion of the planet's atmosphere caused by high extreme UV emissions and strong stellar winds (Lammer *et al.*, 2004). Earth-like planets may lose their atmospheres because of extreme UV-driven hydrodynamic escape if they have a mass less than a specific minimum value that varies with each star (Lammer *et al.*, 2004). This is certainly a phenomenon that deserves to be carefully studied. However, we do not feel that the present state of such investigations precludes the possibility that some M-star planets can retain substantial atmospheres.

If Earth-like planets around M dwarfs do exist and do harbor living organisms, the remaining question is whether we will be able to detect any spectroscopic signature of that life. Those spectral features whose presence or abundance can be attributed to life are called "biomarkers" or "biosignatures" (Des Marais *et al.*, 2002; Kaltenegger *et al.*, 2002). For planets with Earth's atmospheric composition around F, G, and K stars, $O_2$ and $O_3$ are the biosignatures that are most likely to be detectable (Owen, 1980; Leger *et al.*, 1993; Angel and Woolf, 1996; Des Marais *et al.*, 2002; Selsis *et al.*, 2002; Segura *et al.*, 2003). $O_2$ is produced almost entirely by photosynthesis on Earth, while $O_3$ is generated by $O_2$ photochemistry. Nitrous oxide ($N_2O$) and methane ($CH_4$) have been considered as possible biosignatures for Earth-like atmospheres. Biology is believed to be the principal producer of $N_2O$ on modern Earth, with denitrification of the soil by bacteria currently identified as the main source process (Stein and Yuk, 2003). Methane has both biological and non-biological sources. Biological production is by methanogenic bacteria (*i.e.*, methanogens) that live in a variety of anaerobic environments, including the intestines of ruminants (especially cows) and the flooded soils underlying rice paddies. These biological sources appear to outweigh abiotic sources (principally submarine outgassing) by a factor of at least 30. [Kasting and Catling (2003) had estimated this factor to be 300, based on measurements of abiotically produced methane emanating from midocean ridge vents (Kelley *et al.*, 2001). The measured methane concentrations have since increased by a factor of 10 (Kelley *et al.*, 2005), so the ratio of biotic to abiotic production rates is decreased by this same factor. Work in progress demonstrates that the vent methane is indeed abiotic (D. Kelley, personal communication).] Although $N_2O$ has no strong features in the visible below 1.0 µm, it has several bands in the mid-infrared (MIR), principally at 7.8, 8.5, and 17 µm. Methane has relatively strong molecular bands at visible and near-infrared (IR) wavelengths in regions of the spectrum between 0.76 and 1.2 µm, and in the MIR near 7.7 µm. $CH_4$ and $N_2O$ would be hard to detect in the MIR on modern Earth, primarily because of their low concentration (1.7 and 0.3 ppmv, respectively), but also because of the presence of water vapor in our atmosphere, which depresses the observed spectrum strongly in the 5–8 µm wavelength region. However, $CH_4$ might be more detectable on a planet like the early Earth, which had much higher atmospheric $CH_4$ concentrations (Schindler and Kasting, 2000; Segura *et al.*, 2003).

During the next decade a new generation of space-based telescopes, including two Terrestrial Planet Finder (TPF) missions under study by NASA and the Darwin mission being studied by ESA, will hopefully be able to detect and characterize habitable planets around other stars in both the visible/near-IR and thermal-IR parts of the spectrum. The first scheduled NASA mission is TPF-C, which is a coronagraph that will operate in the visible and near-IR. TPF-I and Darwin are envisioned as free-flying interferometers that will operate at thermal IR wavelengths. Despite the large numbers of M dwarfs in the solar neighborhood, these M-star systems are not easy to study with these missions, as their HZs lie extremely close to their parent stars, thus requiring high spatial resolution. The interferometers offer the best possibility for doing so, as their baseline can be lengthened by flying the spacecraft farther apart. However, even TPF-C may be able to observe at least part of the HZ around a few M stars, given its predicted 8-m-long telescope axis and 60 mas (= 4 $\lambda$/D) resolution at 0.6 $\mu$m. A list of possible TPF target stars may be found at http://planetquest.jpl.nasa.gov/Navigator/library/basdtp.pdf.

In this work, we have taken a first step toward understanding which biosignatures may be observable on Earth-like planets around such stars. Using a one-dimensional, coupled photochemical/radiative-convective model we have time-stepped the atmospheric state to equilibrium to determine the abundance of stratospheric ozone and other biogenic gases on hypothetical Earth-like planets around M stars. The M stars chosen as the host stars for this study include two observed M dwarfs (AD Leo and GJ 643, both high-activity M stars) and a suite of hypothetical M dwarfs with different photospheric temperatures, which display no chromospheric activity. These two sets of M stars are used to bracket and illustrate the extremes of the phenomena likely to be observed. As a first approximation we have used time-averaged UV spectra that do not include large flares. A subsequent study will deal with the vastly more complicated, time-dependent problem. Synthetic spectra calculated for these planets will eventually be part of a support library for planet detection and characterization missions, posted at the Virtual Planetary Laboratory website http://vpl.ipac.caltech.edu.

## MODELS AND DATA

*Photochemical and climate models*

A one-dimensional photochemical model (Pavlov and Kasting, 2002) and a radiative-convective model (Pavlov *et al.*, 2000) were used for this work. They were applied in a loosely coupled mode, as described in Segura *et al.* (2003). All of the simulations were performed with a constant surface pressure of 1 atm. The nitrogen ($N_2$) and oxygen ($O_2$) abundances were set equal to Earth's present atmospheric level (PAL), while the $CO_2$ mixing ratio was kept constant at 335 ppmv. The models were tuned to reproduce average observed terrestrial vertical profiles for water vapor, temperature, and ozone (see Fig. 1 in Segura *et al.*, 2003). For the extrasolar planet calculations, stellar fluxes were scaled so as to produce a surface temperature equal to Earth's average surface temperature (~288 K). They were then rescaled to account for the additional greenhouse warming by methane. Thus, the objective of the present study was to see what would happen if Earth were orbiting in the HZ of an M star at the same relative location as for Earth today. The limitations and uncertainties in our codes are discussed by Segura *et al.* (2003).

There are, admittedly, some inconsistencies in the present study. Earth is thought to lie near the inner edge of the Sun's HZ, currently estimated to be located at ~0.95 AU (Kasting *et al.*, 1993). As discussed earlier, a planet farther out in the HZ would need a denser atmosphere and a larger amount of $CO_2$ or some other greenhouse gas to maintain a comparable surface temperature in the presence of lower stellar flux. Unfortunately, such planets cannot be simulated self-consistently with our present model because our long-wavelength (LW) radiative transfer routine, from Mlawer *et al.* (1997), is valid only for a limited range of $CO_2$ partial pressures close to that of modern Earth. [Alternative long-wave radiative transfer algorithms that can handle high $CO_2$ concentrations, *e.g.*, that of Pavlov *et al.* (2000), exist, but they do a poor job in the Doppler region and, hence, yield inaccurate stratospheric temperature profiles.] So, we have chosen to model low-$CO_2$ planets similar to the low-$CO_2$ planets studied previously around F-G-K stars (Segura *et al*., 2003). $CO_2$ has only a small influence on atmospheric photochemistry at 1 PAL $O_2$, so the assumed position of the planet in the HZ should have little influence on our results, other than its direct effect on the strength of the $CO_2$ absorption bands. In high-$CO_2$ atmospheres, the 9.4- and 10.4-μm $CO_2$ bands can either shield or create a possible "false positive" for the 9.6-μm $O_3$ band, as pointed out by Selsis *et al*. (2002). Some sensitivity studies on the effects of higher $CO_2$ levels are described in the Discussion.

We have also restricted our calculations to planets with Earth's present $O_2$ concentration, whereas in our previous work we modeled atmospheres with as little as $10^{-5}$ PAL of $O_2$. The present calculations are intended to illustrate possibilities, as opposed to exhausting them. Low-$O_2$ atmospheres on M-star planets would be better studied with a time-dependent model that captured the dangerously high surface radiation fields that should accompany stellar flares.

The biogenic trace gases $H_2$, $CH_4$, $N_2O$, CO, and $CH_3Cl$ were included in the photochemical model with the same fixed surface fluxes we computed in our previous work (Segura *et al*., 2003). These compounds were originally chosen because they are the more abundant or active biogenic compounds in Earth's present atmosphere. Methyl chloride, for example, is the most abundant chlorine compound and has a direct effect on the tropospheric ozone chemistry. Other biogenic compounds like ammonia or ethylene were not included in the present code because their photochemical lifetimes are short and their spatial distribution is restricted to the troposphere.

The surface fluxes used for the biogenic compounds were those needed to produce their observed mixing ratios in Earth's present atmosphere: $-1.31 \times 10^{12}$ g of $H_2$/year, $9.54 \times 10^{14}$ g of $CH_4$/year, $1.32 \times 10^{13}$ g of $N_2O$/year, $2.35 \times 10^{15}$ g of CO/year, and $7.29 \times 10^{12}$ g of $CH_3Cl$/year. The negative sign on the $H_2$ flux indicates that the calculated flux is downward, *i.e.*, that the surface is a net sink for $H_2$. Most of these fluxes are somewhat higher than more accurate estimates for surface fluxes, which reflects the relatively limited accuracy of our one-dimensional photochemical model. [For example, the best estimate for the modern methane flux is 535 Tg of $CH_4$/year (Houghton *et al*., 1994), or about 40% lower than the value found here.] Using these fluxes is equivalent to assuming that biological production of trace gases is exactly the same as on Earth and that it is *not* affected by such things as changes in the surface UV flux. This might not be realistic, but it is a logical first step in understanding possible biosignatures on M-star planets and arguably more reasonable than assuming fixed mixing ratios, which would imply systematic changes in biogenic production rates for each star examined. Using these assumptions, the two models were iterated until their solutions converged. The resulting pressure and temperature profiles, along with the calculated mixing ratios of $H_2O$, $O_3$, $CO_2$, $CH_4$, $N_2O$, and $CH_3Cl$, were then introduced into a line-by-line radiative transfer model, SMART (the Spectral Mapping Atmospheric Radiative Transfer model). SMART (Meadows and Crisp, 1996; Crisp, 1997) was used to calculate a high-resolution spectrum of each planet for a solar zenith angle of 60°, which approximates the average illumination observed in a planetary disk-average. The radiances shown are the integrated upward fluxes over the hemisphere, divided by π steradians, to approximate the mean radiance seen in the disk-average.

*Stellar data*

Spectra of M stars with complete wavelength coverage are hard to find. Most M dwarfs, even strong dMe flare stars, emit UV fluxes below the limit of detection by space-borne UV telescopes. Flares are easier to observe in the UV, but few, if any, observations have been synchronized to detect UV, visible, and near-IR emission so as to form a complete, contemporaneous spectrum. Theoretical models for M-star photospheric emission are available, but such models lack the chromospheric activity that is responsible for the emission of SW UV radiation ($\lambda < 200$ nm).

For this work, we compiled time-averaged spectra of the stars AD Leo and GJ 643. AD Leo is an M4.5V star, effective temperature 3400 K (Leggett *et al.*, 1996), located at a distance of 4.9 pc from the Sun. This star is one of the most active M dwarfs, and it has been observed over the entire spectral range, albeit during different stages of activity. The International Ultraviolet Explorer (IUE) satellite acquired 114 separate spectra of AD Leo: 64 far-UV spectra (115.1–179.7 nm; SW range) and 50 near-UV spectra (179.7–334.9 nm; LW range). The IUE coverage includes the great flare of 1985 and its aftermath, as well as other epochs in which the presence of strong emission lines signifies intense activity.

To represent the average quiescent spectrum of this star, we avoided all IUE spectra showing obvious emission lines indicative of above-average chromospheric activity, and combined subsets of 40 typical LW and 17 typical SW spectra. We first corrected all these IUE spectra for known artifacts and problems of calibration (Massa *et al.*, 1998) and then reprocessed them using the software described by Massa and Fitzpatrick (2000).

The co-added SW and LW spectra were merged by multiplicative scaling in their region of overlap, preferentially retaining the higher signal-to-noise SW portion. The integrated flux for the AD Leo Lyman $\alpha$ line, $9.97 \times 10^{-12}$ ergs cm$^{-2}$ s$^{-1}$, was provided by Brian E. Wood (University of Colorado, Denver) based on Hubble Space Telescope data. At the top of the planetary atmosphere this flux is $3.9 \times 10^{2}$ ergs cm$^{-2}$ s$^{-1}$. The quiescent optical spectrum from 355.5 nm to 900 nm is from Pettersen and Hawley (1989). It was extended to 1012 nm using an independent spectrum from Leggett *et al.* (1996), who also provide near-IR spectra in the J, H, and K windows covering, respectively, 965.6–1354.5, 1437.5–1853.0, and 1995.2–2409.0 nm. The photospheric emission beyond 2410 nm and into the far-IR was taken from a NextGen model (Hauschildt *et al.*, 1999) atmosphere ($T_{\text{eff}} = 3400$ K, log $g = 5.0$, [Fe/H] = 0.0; see Leggett *et al.*, 1996). The model's synthetic spectrum matches the observed continuum shapes in the JHK windows, though the observed steam bands at the edges of these near-IR windows appear much weaker than in the synthetic spectrum. These discrepancies account for our reliance on observed IR spectra whenever these are available. The resulting complete spectrum was thus composed of empirical IUE data below 335 nm, empirical optical and near-IR spectra from 335 to 2410 nm (with small unobserved regions between the JHK windows), and a NextGen synthetic spectrum longward of 2410 nm.

GJ 643 is an M3.5V star, $T_{\text{eff}} = 3200$ K (Berriman *et al.*, 1992), located 6.5 pc from the Sun. The IUE archive contains 23 spectra. Our UV spectrum was constructed by co-adding all 13 usable SW spectra and five usable LW IUE spectra. In this case, the Lyman $\alpha$ line was included in the IUE observations (see the long line at the left of Fig. 1a) and has a total top-of-atmosphere flux of $2.2 \times 10^{3}$ ergs cm$^{-2}$ s$^{-1}$. The merged IUE spectrum was combined with an observed red optical spectrum (Henry *et al.*, 1994) and with JHK near-IR spectra (Leggett *et al.*, 2000). These data were extended into the far-IR using a NextGen synthetic spectrum for 3100 K, log $g = 5.0$, and [Fe/H] = -0.5 (Leggett *et al.*, 2000).

Because M stars exhibit a large range in activity levels (see Discussion), it is also of interest to examine the detectability of biosignatures for stars with very low levels of chromospheric and coronal activity, the so-called "basal" M stars. Empirical UV spectra are available for only the most active stars, so as extreme examples we have included a subsample of synthetic spectra using purely theoretical model atmosphere spectra, which have no chromospheric emission at all. We selected models with effective temperatures of 3650, 3400, and 3100 K, corresponding to main sequence stars of spectral types M1, M3, and M5, respectively, according to the theoretical effective temperature scale of Allard *et al.* (1997). The spectral type-effective temperature scale remains uncertain for M stars (Leggett *et al.*, 1996; see Jones *et al.*, 2005 and references therein), so the spectral types could be different by one to two spectral subclasses on observed M dwarfs. All of the stars are assumed to have solar metallicity. The non-active model stars were taken from the BaSeL website (www.astro.mat.uc.pt/BaSeL/). The surface Lyman $\alpha$ flux of the M dwarfs is on the order of $10^6$ ergs cm$^{-2}$ s$^{-1}$ (Doyle *et al.*, 1990). This value was scaled for all the planets around model stars, which resulted in a total top-of-atmosphere flux of $2.3 \times 10^{-3}$ ergs cm$^{-2}$ s$^{-1}$ for the M5, $1.6 \times 10^{-3}$ ergs cm$^{-2}$ s$^{-1}$ for the M3, and $1.2 \times 10^{-3}$ ergs cm$^{-2}$ s$^{-1}$ for the M1.

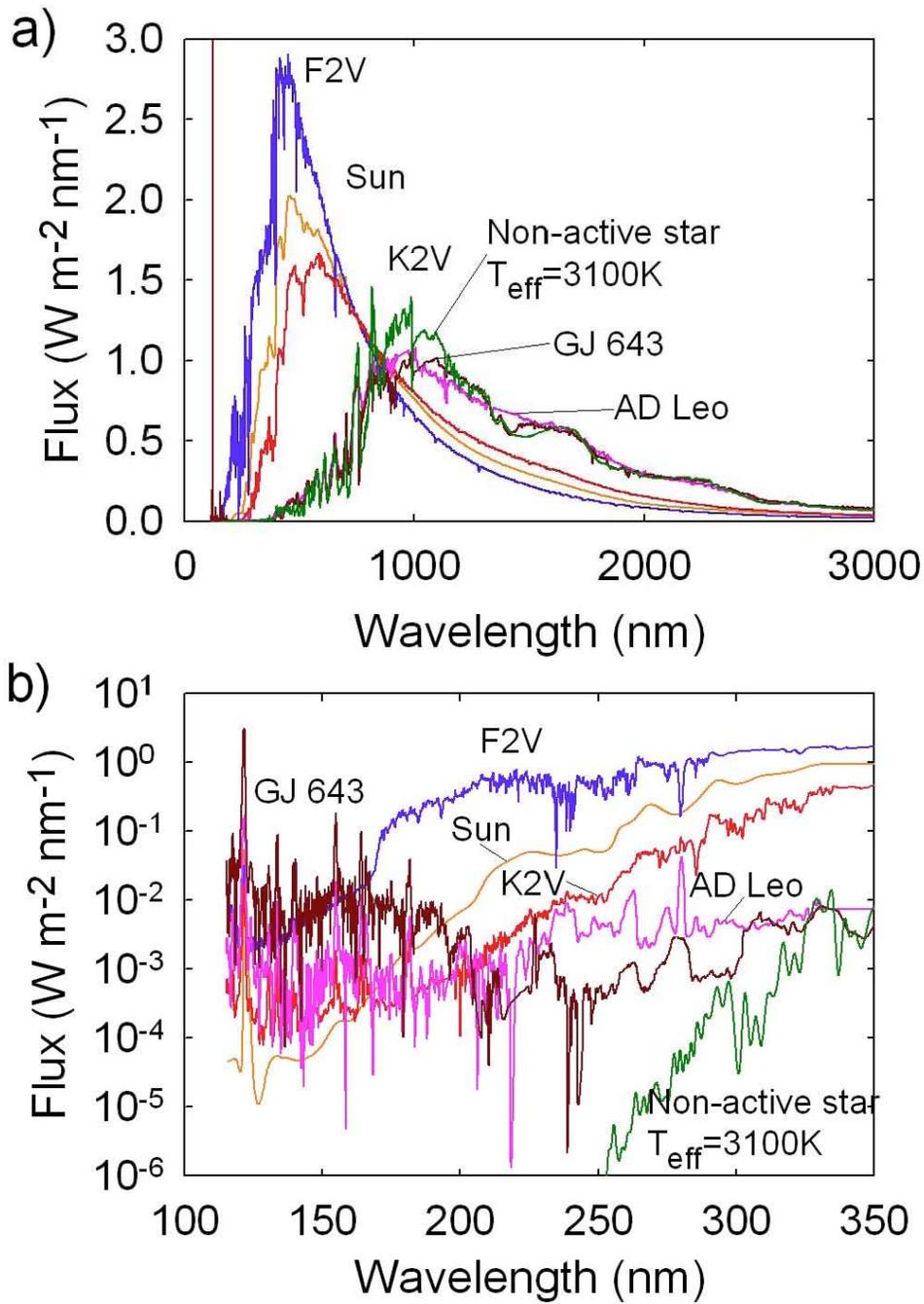

**Figure 1.** Normalized complete (a) and UV spectra (b) for two observed active M dwarfs, a synthetic inactive M star, and for the Sun. Stellar fluxes were normalized in such a way as to produce a surface temperature similar to that on present Earth (288 K).

These models are interpolations in frequency and altitude of the NextGen cool star model atmospheres developed by Allard *et al.* (1997) and should be similar to the Hauschildt *et al.* (1999) model grids. They are notable for their inclusion of tens of millions of molecular and atomic lines in the opacity function. These models compute spectra all the way from the IR down to the far UV, but they treat only the photospheric emission; hence, no chromospheric UV flux is calculated by these models. (This is clear in Fig. 1b: The active M stars exhibit strong UV fluxes, while the model M dwarf emits almost no UV radiation shortward of 250–300 nm.)

## RESULTS

The stellar flux received for each Earth-like planet around one of these stars was calculated by normalizing the full wavelength range stellar spectra so as to produce a mean planetary surface temperature of ~288 K, as described above. As a result of this normalization, we obtained the stellar fluxes at the top of each planet atmosphere, which are shown in Fig. 1. The fluxes used by Segura *et al.* (2003) for the Earth around the Sun, an F2V star, and a K2V star are included for comparison. The normalization takes into account the different albedo that Earth would have for different wavelength distributions of incident radiation (Kasting *et al.*, 1993). For these M stars, the albedo correction ($A_{fac}$) was 0.9, *i.e.*, the normalized flux was reduced by this factor to compensate for the greater absorption of redder stellar radiation by the planet's atmosphere. The final distances of these planets from their parent stars was calculated from $r = 1$ AU $[(L/L_\odot)/A_{fac}]^{1/2}$ and are presented in Table 1.

Figure 1b shows the estimated UV flux from the M stars and from the F-G-K stars studied earlier. According to these data, the two observed M stars outshine the Sun (in a time-averaged sense) at wavelengths shorter than ~190 nm (GJ 643) or ~170 nm (AD Leo). By contrast, for 200 nm $< \lambda <$ 350 nm, the Sun outshines the M stars by a factor of ~100. This is important to the discussion of ozone/methane photochemistry that follows. The non-active (model) M star exhibits almost no UV flux shortward of 250 nm. As the cutoff for $O_2$ photolysis is below this wavelength (Fig. 2), one can expect that a planet around such a star should have a much thinner ozone layer than does Earth. This result was anticipated by W.R. Sheldon in a poster presented at a Gordon Conference during the early 1990s, and the idea was first explored numerically in a paper that he co-authored several years later (Kasting *et al.*, 1997).

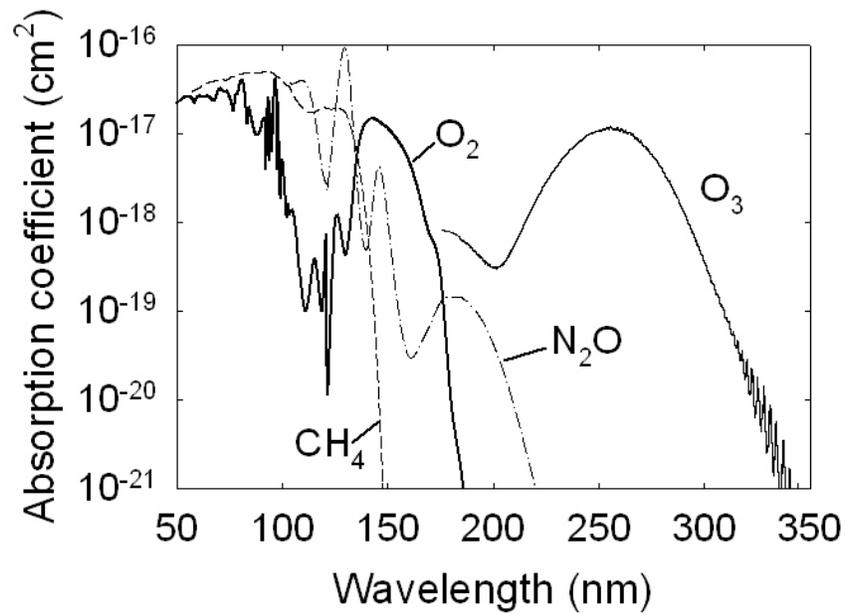

**Figure 2**. Absorption coefficients of O2, O3, N2O and CH4 as a function of wavelength Compiled by Mark Allen (JPL/Caltech). O2: Kirby et al (1979); Watanabe (1985); Ackerman (1971); Kley (1984); Hudson (1974). O3: Denmore et al. (1990). N2O: Denmore et al. (1990). CH4: De Reilhac and Damany (1997); Backx et al. (1975); Lee et al. (1997); Lee and Chiang (1983); Mount et al. (1997); Mount and Mood (1978).

Spectral signatures from planets are produced in the visible via interactions of the planetary atmosphere and surface with the reflected radiation from the parent star. In the MIR, these features are produced when emitted surface and atmospheric thermal radiation interacts with the atmosphere as it escapes to space. Consequently, at visible wavelengths the detectability of a spectral feature is relatively insensitive to temperature and influenced primarily by the column-integrated abundance of a given constituent in the atmosphere. However, in the case of the MIR, the observed features depend not only on the concentrations and vertical distribution of the components of the planet's atmosphere, but on the vertical temperature profile as well. Figure 3 shows the temperature, ozone, and water vapor profiles of Earth-like planets circling the active M stars and the Sun. The results of the non-active M-star planets are not shown on these graphics because their atmospheres are not able to reach equilibrium given the biogenic fluxes we have used in the model. The reason for this behavior will be explained later in this section.

Although the active M-star planets in our sample exhibit ozone layers that are comparable to that on Earth (Table 2 and Fig. 3b), their stratospheres are cooler than that of present Earth (Fig. 3a). This is due primarily to the lower near-UV fluxes received by the M-star planets. Indeed, the planet circling GJ 643 has an ozone layer that is 50% thicker than Earth's. This result is expected, as both stars emit much UV radiation shortward of 200 nm, which is capable of dissociating $O_2$, thereby forming $O_3$. The high water vapor abundances in the upper stratospheres of the active M-star planets (Fig. 3c) are a consequence of their high methane concentrations. Methane is oxidized in the stratosphere by the (net) reaction: $CH_4 + 2\ O_2 \rightarrow CO_2 + 2\ H_2O$. Thus, each $CH_4$ molecule that is oxidized results in the production of $2\ H_2O$ molecules. Despite the high water vapor abundances in the stratosphere, $H_2O$ remains well above condensation values in all cases, so stratospheric clouds should not be present.

In Earth's present atmosphere, the biogenic compounds methyl chloride ($CH_3Cl$), $N_2O$, and $CH_4$ are present in small concentrations (0.5 ppbv, 0.3 ppmv, and 1.7 ppmv, respectively). However, on planets around M stars, their concentrations could conceivably be much larger. Other biogenic trace gases might be more abundant as well, but these are the only ones that we have studied. Figure 4 shows calculated vertical mixing ratio profiles for $N_2O$, $CH_4$, and $CH_3Cl$ for Earth and for the active M-star planets. The results for $N_2O$ are what one might expect, given that this gas photolyzes at wavelengths shorter than 220 nm (Fig. 2). The $N_2O$ concentration for the active M-star planets increases over the Earth's current concentration by a factor of 3, which puts it at ~1 ppmv. $N_2O$ concentrations of this magnitude will likely still be difficult to detect with missions like TPF or Darwin, except for stars in which the UV chromospheric emission is far below the levels in strong flare stars. As explained in the next section, it is possible that most M dwarfs fall into this category.

The results for $CH_4$ are surprising: The active M-star planets have concentrations exceeding 300 ppmv—an increase by nearly a factor of 200 relative to present Earth. This behavior can be explained by the effect that the M star spectral energy distribution has on the photochemical removal of $CH_4$. In high-$O_2$ atmospheres like Earth's, photolysis of methane (requiring $\lambda < 145$ nm) is slow, so the high far-UV flux from M stars has little effect on its abundance. Instead, methane is destroyed mainly by the same photochemical reaction sequence that is important in Earth's troposphere today:

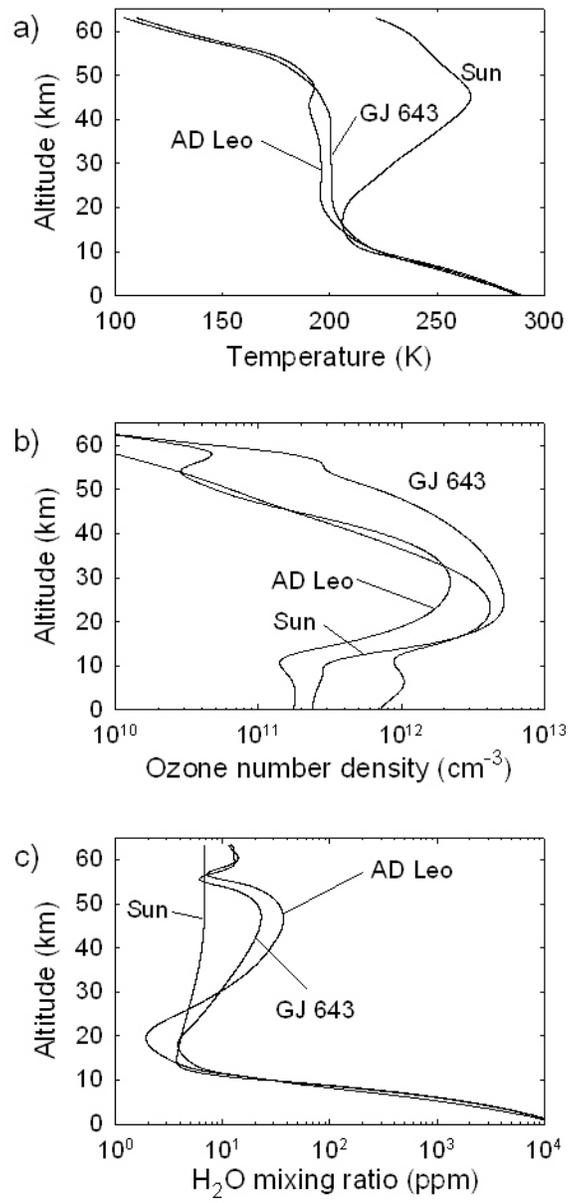

**Figure 3**. Model results for a planet like Earth around two active M dwarfs and the Sun: (a) temperature, (b) ozone number densities, and (c) water mixing ratios.

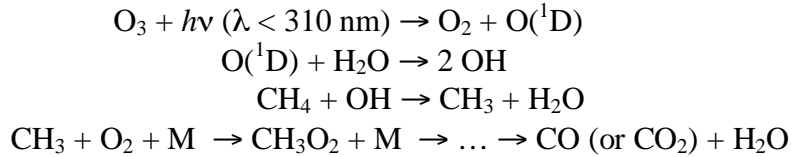

$$O_3 + h\nu \ (\lambda < 310 \ nm) \rightarrow O_2 + O(^1D)$$
$$O(^1D) + H_2O \rightarrow 2 \ OH$$
$$CH_4 + OH \rightarrow CH_3 + H_2O$$
$$CH_3 + O_2 + M \ \rightarrow CH_3O_2 + M \ \rightarrow \ldots \rightarrow CO \ (or \ CO_2) + H_2O$$

This reaction sequence takes place throughout the troposphere and lower stratosphere. Most of the photons that create $O(^1D)$ in Earth's lower atmosphere are at wavelengths between 200 nm and 310 nm, as $O_2$ absorbs wavelengths shorter than 200 nm high up in the stratosphere or mesosphere.

As we saw earlier, M dwarfs emit very little UV radiation at wavelengths between 200 nm and 300 nm (see Fig. 1b). Active M stars emit lots of shorter-wavelength UV radiation, but these photons are used primarily to dissociate $O_2$, as both its absorption coefficient and concentration are higher than those of other atmospheric gases (Figs. 2 and 4). Consequently, little $O(^1D)$ is produced in the atmospheres of our M-dwarf planets, and the predicted tropospheric concentrations of the hydroxyl radical, $OH^\bullet$, are lower by a factor of $10^5$ than they are on Earth (~1 cm$^{-3}$ compared with ~$10^6$ cm$^{-3}$). Thus, instead of having a photochemical lifetime of 10–12 years (Houghton *et al.*, 1994), as on Earth, $CH_4$ has a predicted lifetime of about 200 years on an M-star planet. The lifetime of methane is *not* simply inversely proportional to [OH], as methane can diffuse upward through the atmosphere to altitudes where more UV photons and more OH are available. This is calculated, though, in our photochemical model. In the troposphere, OH is the major sink of methyl chloride, as well. Because of the reduced concentration of OH on the M-star planets, the concentration of $CH_3Cl$ increases dramatically compared with the terrestrial value (Fig. 4c).

The synthetic planetary spectra for Earth-like planets in orbit around M stars clearly show the effect of the different stellar spectral energy distributions. This is apparent in both the visible and MIR spectra. For the planet orbiting AD Leo [and GJ 643(data not shown)], ozone and methane have strong absorption features in the MIR (9.6 μm and 7.7 μm, respectively) (Figs. 5 and 6), with the $O_3$ feature being of comparable strength to Earth's. In contrast, the $CH_4$ feature is far stronger for the AD Leo planet than for Earth, as a consequence of the much higher $CH_4$ concentration. Interestingly, methyl chloride, which has spectral features in the MIR in the range 6.5–7.5, 9.3–10.3, and 13–14.8 μm, is also seen in the AD Leo planet spectrum, where it is relatively easy to detect, compared with the Earth spectrum.

Figure 7 shows the comparison of the Earth and the AD Leo planet with and without methyl chloride to illustrate the effect of methyl chloride in the spectra of the AD Leo planet. It can be seen that the $CH_3Cl$ feature at 9.3–10.3 μm coincides with the $O_3$ 9.6-μm band, and acts to enhance and broaden the absorption feature associated with the $O_3$ band at 9.6 μm. Methyl chloride absorption is also seen strongly in the 13–15 μm region, and a weaker feature can be discerned near 7 μm. The $CH_3Cl$ modification of the strength of the 9.6-μm ozone band could be mistakenly used to infer higher $O_2$ levels. To guard against this misinterpretation, it would be necessary to look for the 13–15-μm $CH_3Cl$ absorption feature as well. However, the 13–15-μm feature also extensively modifies the SW profile of the $CO_2$ band, and could introduce errors in the interpretation of the atmospheric temperature profile using the $CO_2$ band. SEGURA ET AL: Planets around M stars

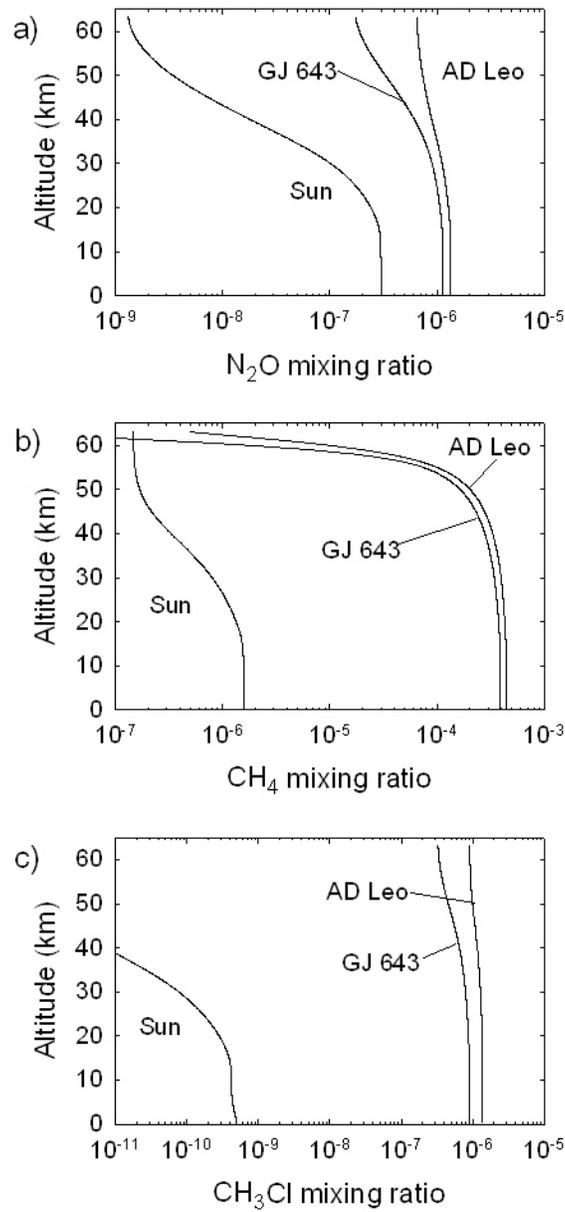

**Figure 4.** Vertical mixing ratios of the biogenic gases (a) N2O, (b) CH4 and (c) CH3Cl for Earth-like planets around two M dwarfs and for Earth itself. Planets are identified by their parent stars. For Earth, the surface mixing ratios of the gases are fixed at their observed values. For the M-star planets, the surface flux is fixed at the value calculated for Earth.

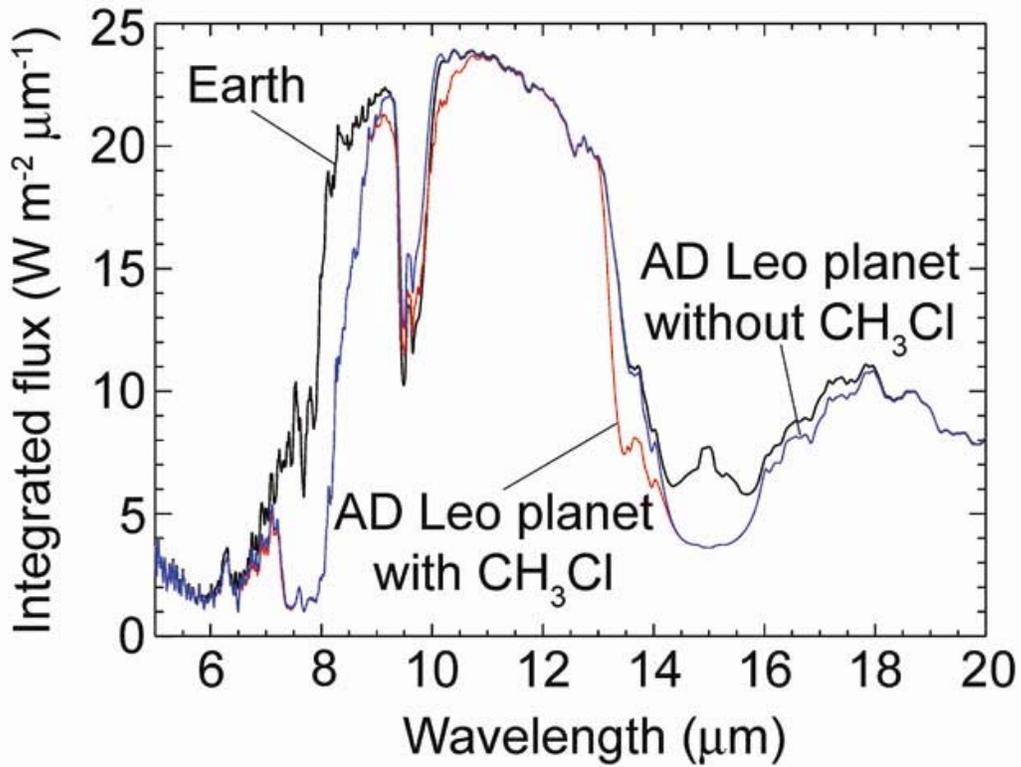

**Figure 7**. IR spectra from an Earth-like planet around AD Leo with CH3Cl (red) and without CH3Cl (blue). The spectrum from Earth is shown for comparison (black).

On Earth, 60% of the sources of $CH_3Cl$ are unknown, while the known sources come from biological and human activity. These include tropical forests (Yokuchi *et al.*, 2002), biomass burning, planktonic algae (Harper *et al.*, 2003), and degradation of organic matter in soils (Keppler *et al.*, 2000). Other human activities, such as the incineration of municipal and industrial wastes, are relatively minor sources (<10% of the known sources).

$CO_2$ is not a biosignature, but it is the most prominent indicator of a terrestrial planet atmosphere that indicates potential habitability. $CO_2$ can also be useful for probing the temperature profile of the planetary atmosphere, even at low spectral resolution. For example, near the center of the 15-$\mu$m $CO_2$ band, the Earth's spectrum shows an emission peak, which indicates a hot stratosphere. By contrast, the M-star planets have deep, flat, 15-$\mu$m $CO_2$ bands—a result of their cooler, almost isothermal stratospheres. This discrepancy in stratospheric temperature can be seen most readily in the brightness temperature spectrum (Fig. 5b), from which it is possible to infer that the AD Leo planet's stratosphere is some 50 K cooler than Earth's. Note also that the water vapor absorption from 5–7 $\mu$m and 19–25 $\mu$m is of comparable intensity in all three cases, despite the large differences in stratospheric water abundance. At these wavelengths, the absorption bands reflect the $H_2O$ abundance and temperature structure of the troposphere, which are very similar in all three cases.

The synthetic spectra of these planets in reflected visible light show marked differences from the spectrum of Earth (Fig. 8a). Most of this difference is caused by the highly dissimilar spectra of the starlight itself, which includes strong molecular absorption in the M-star atmospheres. To study the planet at visible wavelengths, it is necessary to divide the planet's spectrum by the incident stellar flux at the top of the planet's atmosphere. This produces a reflectivity, or albedo, spectrum, as shown in Fig. 8b. Two major differences are seen between the reflected spectrum of M-star planets and that of Earth. The most obvious of these is the presence of stronger $CH_4$ absorption throughout the wavelength range 0.84–1.25 $\mu$m, especially between 1.1 $\mu$m and 1.2 $\mu$m. The second difference can be seen in the Chappuis bands of $O_3$ between 0.5 $\mu$m and ~0.7 $\mu$m, where the planet with the thicker $O_3$ layer, Earth, clearly has the most absorption. Note that the 0.76-$\mu$m $O_2$ A-band is of equal intensity in both cases. It should be borne in mind that other biogenic trace gases might also build up to high concentrations in an M-star planet atmosphere, so the actual spectrum of the planet might be even more complex than shown here. Regardless, these models provide a basis for interpretation of observational spectra from M-star planets.

As mentioned earlier, we also simulated a set of Earth-like planets around quiescent M dwarfs. The low flux of these stars at all UV wavelengths produces drastic effects on the atmospheric photochemistry of the simulated planets. As explained above, the main sink of methane is its reaction with OH, which is produced indirectly from ozone photolysis. On the planets around our quiescent stars, ozone photolysis is extremely slow, and tropospheric OH densities drop to ~$10^{-2}$ molecules/cm$^3$. Consequently, the lifetime of $CH_4$ is extremely long. Indeed, the destruction of methane is so slow that it is unable to keep up with a biological production rate equal to that of modern Earth. This leads to a phenomenon that might be termed "methane runaway," in which $CH_4$ concentrations increase without bounds in our photochemical model. This, of course, is not physically realistic. In reality, some physical process (*e.g.*, escape to space) would limit the $CH_4$ concentration, or it would become thermodynamically unprofitable for the biota to continue to produce $CH_4$.

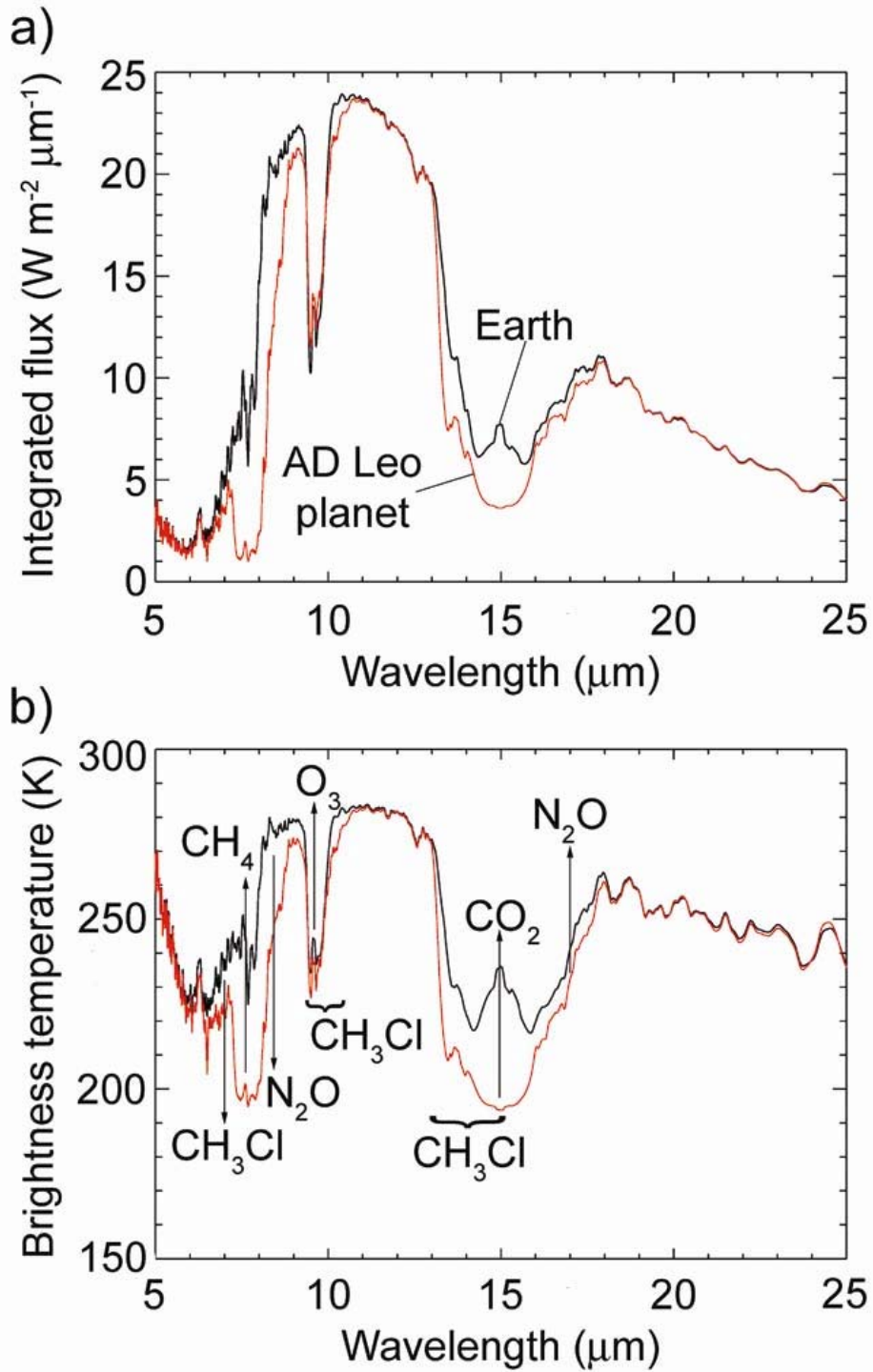

**Figure 5. a)** Complete IR spectra (5 to 25 µm) for Earth-like planets around the Sun (black) and AD Leo (red). **b)** Brightness spectra for the same planets.

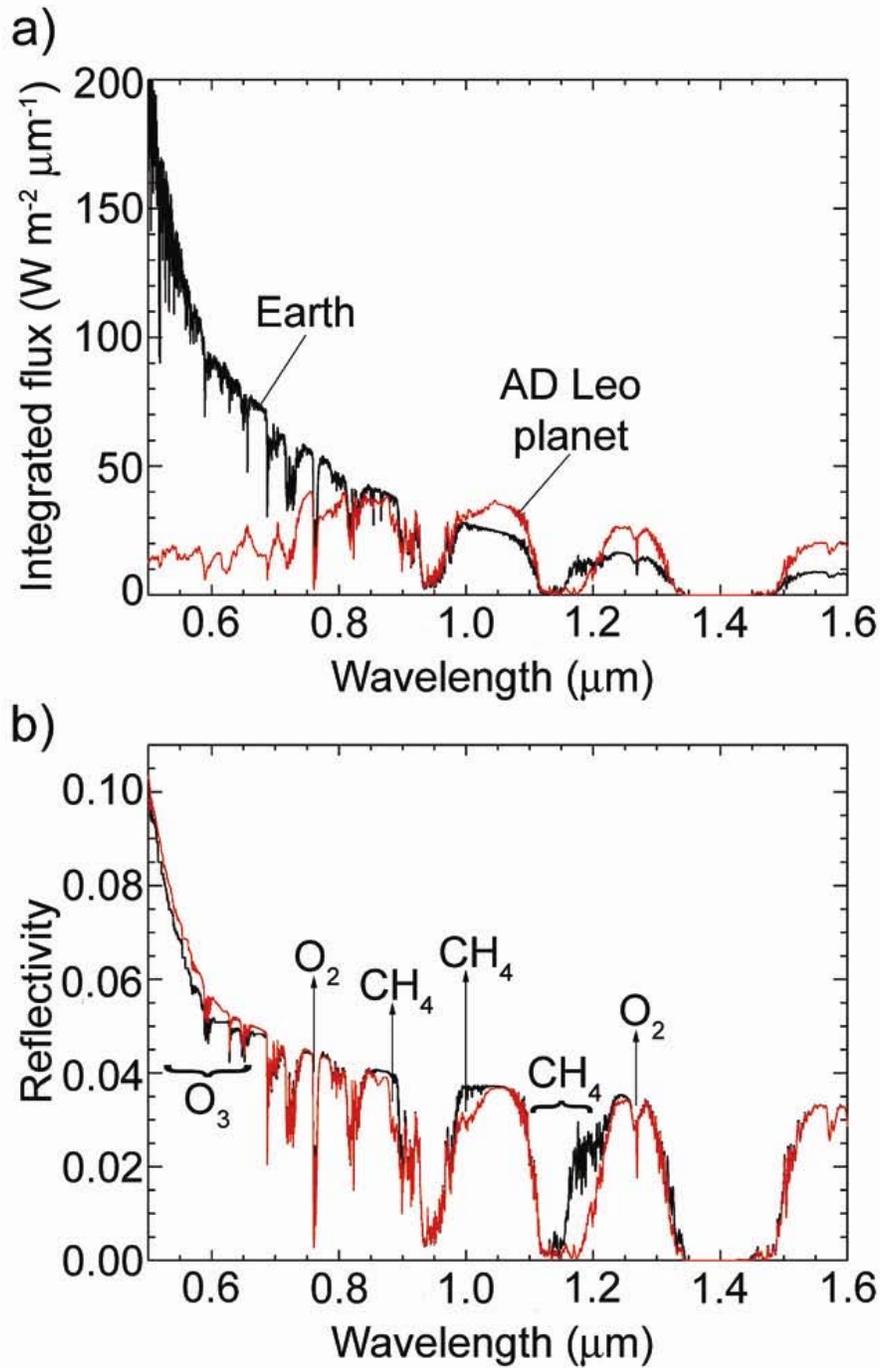

**Figure 8**. Complete visible/near IR spectra (0.5-1.5 µm): **a)** Spectra for Earth-like planets around the Sun (black), AD Leo (red) **b)** Reflected spectra for the same planets. Black--Earth; red--AD Leo planet.

To get around this problem, we changed the lower boundary condition for methane in the photochemical model from a fixed flux to a fixed mixing ratio of $5 \times 10^{-4}$ (500 ppmv). This value is approximately what would be maintained if volcanoes produced the same amount of reduced gases as on Earth and hydrogen escaped with the diffusion-limited rate (Hunten, 1973; Walker, 1977; Kasting and Catling, 2003). (This would entail some high-altitude breakdown of $CH_4$ into $H_2$ or H that is not explicitly included in the model.) We also changed the lower boundary conditions for $H_2$ and CO to fixed deposition velocities of $2.4 \times 10^{-4}$ cm/s and $1.2 \times 10^{-4}$ cm/s, respectively. These values correspond to maximum air–sea transfer rates estimated using the "piston velocity" approach (Broecker and Peng, 1982; P. Kharecha et al., manuscript in preparation). (If this assumption is not made, then $H_2$ and CO can also "run away" in these model atmospheres.) Given these boundary conditions, the methane flux necessary to sustain 500 ppmv of methane is $9.4 \times 10^{14}$ g/year (99% of our present Earth value) for the planet around the hottest star ($T_{eff}$ = 3650 K), $5.5 \times 10^{14}$ g/year (57% of the Earth value) for the planet around the star with $T_{eff}$ = 3400 K, and $2 \times 10^{14}$ g/year (21% of the Earth value) for the coolest star of our sample ($T_{eff}$ = 3100 K). High mixing ratios of $N_2O$ and $CH_3Cl$ were also obtained for all the quiescent stars (Fig. 9). Non-active stars have less $CH_3Cl$ than active stars because the OH abundance on the quiescent M stars ranges from $10^2$ to $10^3$ molecules/cm$^3$, while the active stars have ~1 molecule of OH/cm$^3$.

The intriguing implication of these results is that the simultaneous detection of $O_2$ (or $O_3$) and $CH_4$, $N_2O$, or $CH_3Cl$ in a distant planetary atmosphere may be significantly easier for Earth-like planets orbiting M stars than for an Earth-like planet around a star like our Sun. This detection would be most easily accomplished in the thermal-IR, where the strong 7.7-μm $CH_4$ band and 9.6-μm $O_3$ band are clearly visible in our simulated spectra (Fig. 6). The simultaneous detection of $O_2$ and methane might also be possible in the visible/near-IR as well using the 0.76-μm $O_2$ band and either the 1.0-μm or the stronger, and likely more detectable, 1.2-μm $CH_4$ band. Lovelock (1965) suggested many years ago that the simultaneous presence of $O_2$ and a reduced gas such as $N_2O$ or $CH_4$ would constitute the strongest possible spectroscopic evidence for life. [Lederberg (1965) had made a similar argument about thermodynamic disequilibrium as evidence for life without specifying the nature of the compounds to be observed.] Thus, despite the caveats to M-star planet habitability raised in the Introduction, the potential enhanced detectability of these atmospheric biosignatures for planets around M stars may prompt architects of future planet detection and characterization missions to include M stars as targets.

As mentioned previously, a planet near the outer edge of the HZ would require more $CO_2$ to remain habitable. To test the effect of such a change, we repeated our photochemical calculations using a $CO_2$ concentration of 3.35% (100 PAL). (We did not do a coupled climate model calculation, as we are assuming that this $CO_2$ increase would be accompanied by a decrease in the incident stellar flux.) The result for the AD Leo planet was to increase the amount of methane and methyl chloride in the planet's atmosphere by ~30%. This may happen because the excess CO produced from $CO_2$ photolysis reacts with OH, which makes the latter species even less abundant than before (and, hence, increase the lifetime of methane and methyl chloride). The concentration of $N_2O$ is not affected by the increase of $CO_2$. Surprisingly, the same calculation for the Earth itself, and for an Earth-like planet around a K2V star (epsilon Eridani), yields a small (<10%) decrease in the methane mixing ratio. In these cases, OH is almost unaltered by the increase in $CO_2$. (Recall that these planets receive much more 200–300 nm radiation, so their tropospheric chemistry is entirely different.) We conclude that changes in $CO_2$ concentration would only increase the potential for simultaneous observation of $CH_3Cl$, $CH_4$, and $O_2$.

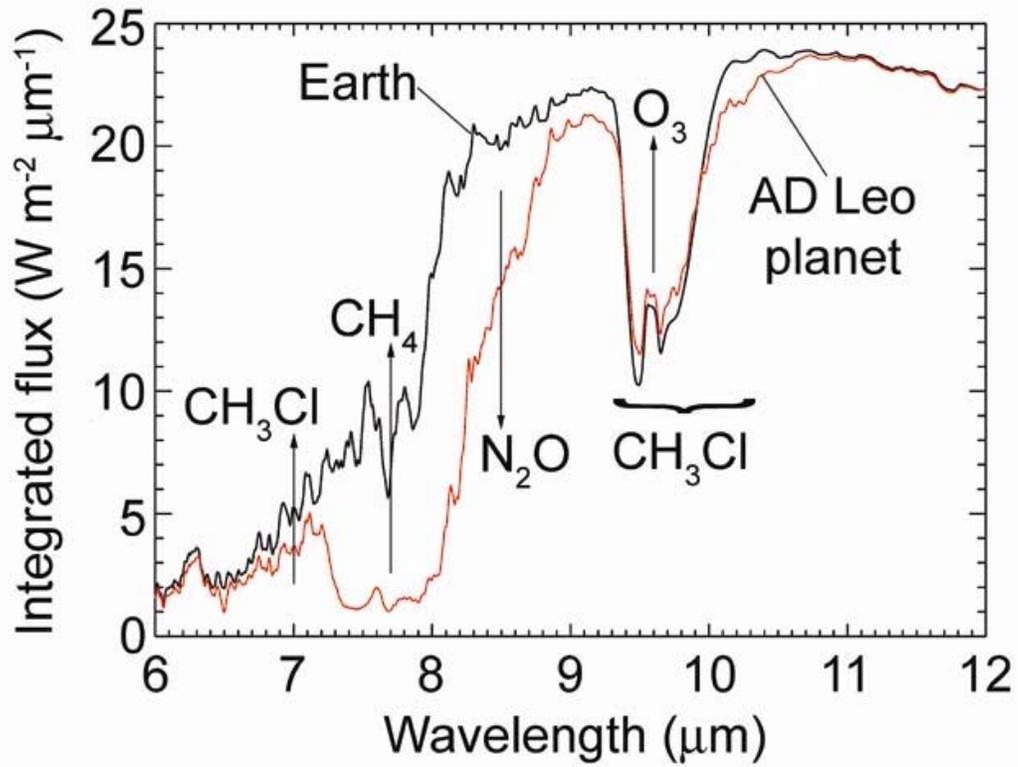

**Figure 6**. Close-up IR spectra from 6 µm to 12 µm: Black--Earth; red--AD Leo planet.

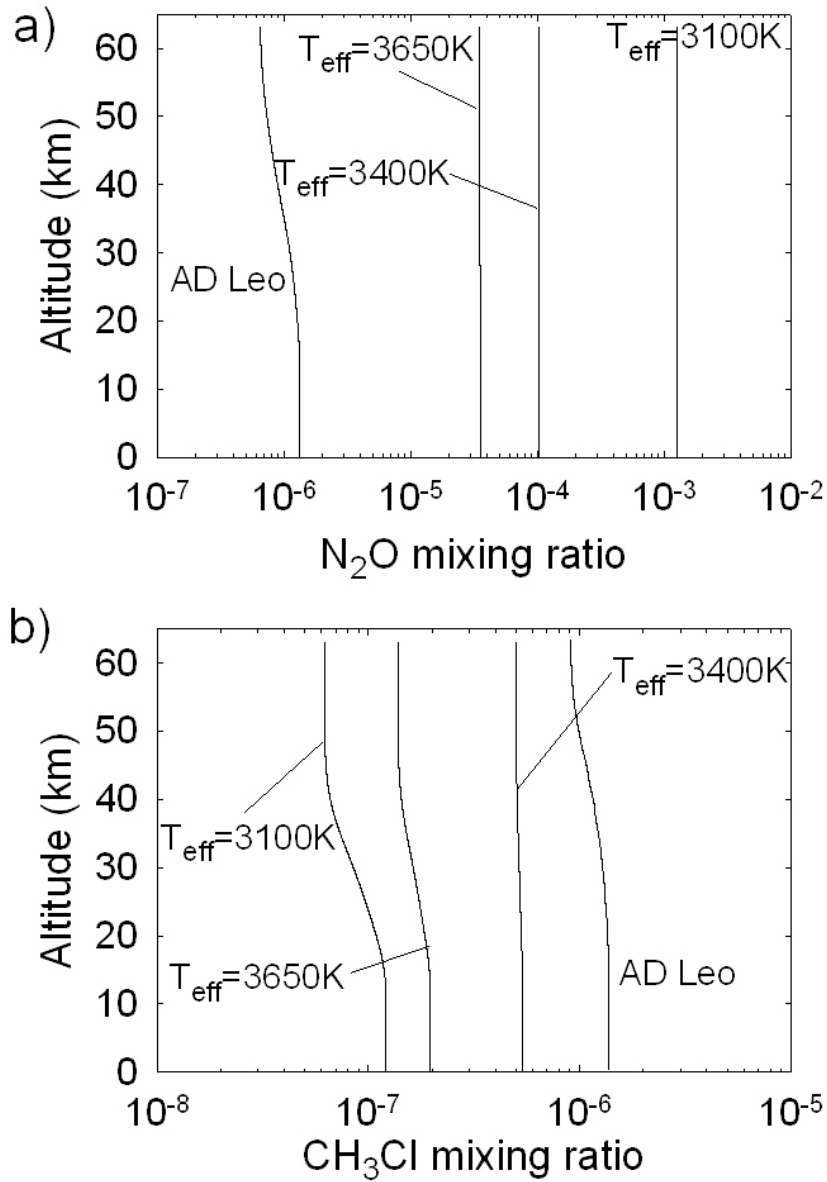

**Figure 9**. Vertical mixing ratios of the biogenic gases (a) N2O and (b) CH3Cl for Earth-like planets around quiescent M dwarfs and for Earth itself. Planets are identified by the effective temperature of their parent stars. For Earth, the surface mixing ratios of the gases are fixed at their observed values. For the quiescent M-star planets the boundary conditions were: fixed CH4 mixing ratio (500 ppmv), fixed deposition velocities for H2 ($2.4 \bullet 10^{-4}$ cm/s) and CO ($1.2 \times 10^{-4}$ cm/s) and fixed surface flux for methyl chloride and nitrous oxide.

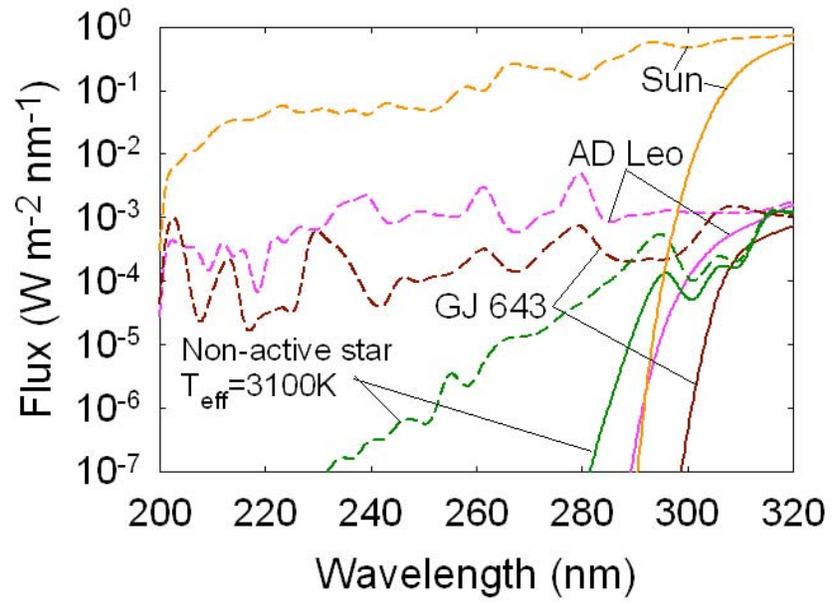

**Figure 10.** Incident (dashed lines) and surface (solid lines) UV fluxes for Earth-like planets around M dwarfs and around the Sun.

*Surface UV fluxes on Earth-like planets around M dwarfs*

The presence of ozone in a planetary atmosphere is not only relevant to the remote detection of life, but also provides a shield against UV radiation that can damage the cells and DNA of organisms. Our photochemical model predicts surface UV fluxes and, hence, can be used to estimate biological damage. For this purpose, the near-UV spectrum is traditionally divided into three ranges. On Earth's surface, we are well protected from radiation at wavelengths shorter than 280 nm (known as UV-C); therefore, most of the research has been done in the UV-B (280–315 nm) and UV-A (315–400 nm) ranges.

As shown in our previous work (Segura *et al.*, 2003), a thick ozone layer can prevent harmful UV radiation from reaching the surface of a planet even when the stellar flux in this spectral region is an order of magnitude larger than that of the Sun. Hence, it is not surprising that on the planets around AD Leo and GJ 643 the ozone layer blocks most of the incident UV radiation (Table 3 and Fig. 10). In contrast, the quiescent M-star planets have ozone column densities less than 40% of that of Earth (Table 2). This is enough to effectively shield the surface of the planet from UV. The surface UV-B flux for planets around these stars is less than 2% of the Earth value. For the whole UV range (200–400 nm), the surface UV flux is 20 times lower than the UV flux that reaches the Earth surface (Table 3).

The surface UV fluxes alone are not a direct measure of the magnitude and type of damage that this radiation can cause to living organisms. To estimate this, it is necessary to use action spectra that describe the relative effectiveness of different wavelengths in inducing damage. UV dose rates were calculated by convolving the surface UV fluxes with an action spectrum for DNA damage (Van Baalen and O'Donnel, 1972). We realize that action spectra vary depending on the type of DNA lesion (*e.g.*, dimer production vs. double strand breaks) and environmental conditions. [See Jagger (1985) for action spectra and Sinha and Hader (2002) for a concise review of UV-induced DNA damage and various repair pathways.] Thus, our calculated surface biological doses should be considered illustrative. Use of other action spectra to assess the UV damage and mutation rates on habitable planets is deferred to future work. The results were normalized to the damage received on Earth (Table 4). M-star planets would be safe places to live from a UV standpoint. One might wonder, though, whether the lower UV dose rates might slow mutation rates and, hence, hinder biological evolution.

## DISCUSSION AND EXTENSIONS

*M dwarf activity levels and implications for possible biosignatures*

The theoretical model M-star synthetic spectra represent an absolute minimum (zero) in terms of chromospheric UV emission. However, the two actual stars whose spectra we have used, AD Leo and GJ 643, may represent an upper limit, as the reason UV spectra are available for them is precisely because of their extremely high (AD Leo) or high (GJ 643) activity levels. The distribution of M stars between these two extremes is obviously important in assessing the applicability of our results to specific target stars in the future. Although we do not have UV fluxes for many M stars, we can examine the distributions of chromospheric and coronal proxies for near-UV chromospheric emission, namely, Hα and CaII H & K emission in the visible, MgII h & k emission in the near UV, and soft x-ray emission.

First, we consider the fraction of M stars that are chromospherically active according to some criteria. West *et al.* (2004) used a large sample of nearly 8,000 M0 to L0 stars from the Sloan Digital Sky Survey to quantify the activity level, as measured by the ratio of Hα to bolometric luminosity, as a function of spectral type. With their criteria for activity (*e.g.*, the Hα equivalent width must be >1.0 Å), they found that only about 24% of these stars were classified as active, with the active fraction increasing from near zero at M0 to 50% at M5, and peaking at about 75% at M8. These results evidently do not depend strongly on the sensitivity limit, as demonstrated by the similar results obtained by Mohanty and Basri (2003), who studied a much smaller sample of stars later than M4 with a sensitivity limit of about 0.2 Å for Hα equivalent width.

We also considered various measures of spread among activity indicators, for active star samples and for stars with the weakest activity indicators. West *et al.* (2004) found that $L_{H_\alpha}/L_{bol}$ varied by over two orders of magnitude among the active stars, with the variation being about an order of magnitude at a given spectral type. A similarly large spread was found by Mohanty and Basri (2003) for their sample of M4 to L6 dwarfs. Considering the large fraction of inactive stars (which will have smaller but unknown values of $L_{H_\alpha}/L_{bol}$) and the spread among the active stars, this measure of chromospheric emission could vary by at least three orders of magnitude among all M stars. The same could be true of the UV flux in the HZ (since $L_{H_\alpha}/L_{bol}$ and $L_{UV}/L_{bol}$ do not depend on distance from the star), assuming that Hα and UV fluxes are correlated.

The soft x-ray luminosity measured by the ROSAT satellite is a measure of coronal activity and known to be correlated with chromospheric activity indicators. The ratio $L_X/L_{bol}$ given for over a hundred K and M stars within 7 pc by Fleming *et al.* (1998) shows a range of over three orders of magnitude. An enlargement of the sample to include several hundred K-M stars using Hipparcos distances and a somewhat revised bolometric correction scale gives a similar result (J.S., unpublished data). Indeed, the average value of $L_X/L_{bol}$ for about a hundred M stars in common with the UV Ceti (flare, or emission) star catalogue of Gershberg *et al.* (1999) is nearly two orders of magnitude larger than the average ratio for non-active M stars. This result is consistent with a number of specific cases, such as the values of stellar surface x-ray fluxes and other activity indicators for AD Leo and two relatively inactive stars (Gl 588 and Gl 628) given by Mauas *et al.* (1997).

The study of IUE Mg II h & k line (~280 nm) fluxes in a sample of K and M stars by Mathioudakis and Doyle (1992) showed a spread of two orders of magnitude in this flux, even among active dMe stars. We have not carried out a comparison of M stars detected in the four bands of the Extreme Ultraviolet Explorer (EUVE) satellite, whose centroid wavelengths vary from 20 to 62 nm (Bowyer *et al.*, 1994), as only the most active stars were detected. (See the large upper limits for candidate low-activity stars given by Mathioudakis *et al.*, 1994.)

We can also give a more specific and quantitative estimate of the expected range by comparing the stars with the largest and smallest activity. Main sequence stars with the lowest, or undetectable, levels of activity are usually referred to as "basal stars," and the result that there is a lower limit to chromospheric activity at a given color has been known for decades (*e.g.*, Wilson, 1968; Cram and Mullan, 1979). The basal level is often attributed to residual heating by acoustic dissipation in stars whose magnetic activity is very low. We have taken Ca II, Mg II fluxes for candidate basal stars from Mathioudakis *et al.* (1994), Doyle *et al.* (1994), and Mauas *et al.* (1997), and $L_X/L_{bol}$ for these same stars from Fleming *et al.* (1995). We compared the average values with the corresponding average values for the flare stars AD Leo, AX Mic, AU Mic, EQ Peg, and YZ CMi given by Houdebine *et al.* (1996), adopting AD Leo data as given by Mauas *et al.* (1997). (We omit BY Dra, as its activity is due to its close binary nature, and Proxima Cen because it is a relatively weak active star, as measured by flare amplitudes, x-ray flux, or any chromospheric indicator.)

For the basal stars, lower limits were included in the averages. We find that the average Ca II and Mg II fluxes (in erg cm$^{-2}$ s$^{-1}$) are $1.6 \times 10^4$ and $3.4 \times 10^4$, respectively, for the candidate basal stars, while for the flare stars the average fluxes are $7.9 \times 10^5$ and $9.2 \times 10^5$. For the average values of log ($L_X/L_{bol}$) we find −5.0 for the three candidate basal stars with available data and −3.7 for four of the flare stars; the latter number becomes −3.1 if the weak flare star AX Mic is excluded from the sample. By comparison, the nearest M star, Proxima Cen, has chromospheric fluxes intermediate between the basal stars and strong flare stars, and very small log ($L_X/L_{bol}$) = −5.5, while AD Leo is near the maximum for all activity indicators.

We conclude that, although observations in the 200–300 nm region are not available for any but the few most active M stars, Hα (visual), X-ray, Ca II (visual), and Mg II (UV) proxies for the UV chromospheric continuum all suggest that a range in overall flux level of two to three orders of magnitude should be expected in this wavelength range, with AD Leo near the maximum observed. This variation should be borne in mind when considering those of our results that depend sensitively on the UV flux (*e.g.*, ozone column depth, $N_2O$ band strength, surface UV-C flux, and surface UV doses for biological damage). We are hopeful that future UV satellite observations will provide near-UV spectra for nearby less-active stars. Unfortunately, the known candidate basal stars are more distant than about 8 pc. The nearest star, Proxima Cen, which is intermediate between basal and highly active stars, seems like a promising target.

*Impact of M dwarf variability on possible biosignatures*

The results found here are promising in that they suggest that, with sufficient wavelength coverage, Earth-like biosignatures may be as detectable, if not more detectable on habitable planets around M stars, than for the Earth itself. But, the presence and detectability of these biosignatures could be affected by the time variability of these host stars. As mentioned earlier, flare activity on M stars can last from a few seconds to minutes, or even hours. For AD Leo, a flare with an energy of $3 \times 10^{32}$ ergs in the B-band (centered at 440 nm) occurs every ~4 days (Fig. 5 in Gershberg, 1989). By comparison, solar flares with this frequency emit no more than $10^{29}$ ergs on the same band. During the great AD Leo flare observed in 1985 (Hawley and Pettersen, 1991), $7.1 \times 10^{32}$ ergs were emitted in the 120–326 nm wavelength range in an impulsive phase that lasted 25 min (Table 6 in Hawley and Pettersen, 1991). This implies that an Earth-like planet in the HZ would have received an additional 30.3 W/m$^2$ during this flare stage, or about 1/10$^{th}$ of the average total stellar output. The impulsive phase of this flare was followed by a gradual phase that lasted almost 4 h. During this period, the planet would have received an additional 0.25 W/m$^2$ in this wavelength range. A planetary atmosphere subjected to these flux variations with these intensities may never reach a true steady state. Without performing such a simulation, it is unclear how this time variability would affect the abundances of biosignature gases such as $O_3$ and $CH_4$.

## CONCLUSIONS

We have calculated atmospheric biosignature concentrations, and generated synthetic visible/near-IR and thermal-IR spectra for Earth-like planets circling active and quiescent M dwarfs. The atmospheres of the quiescent M-star planets may not be stable if methane or carbon monoxide fluxes are similar to those on Earth because the photochemical sink for these compounds is much smaller because of the low incident UV flux. Predicted concentrations of methane, nitrous oxide, and methyl chloride are higher for the M-star planets than for Earth, and these compounds are more detectable on those planets. The planets around the active M stars developed ozone layers similar to that on Earth. Thus, the signature of $O_3$ from habitable planets around active M dwarfs, along with the signatures of various reduced gases, may be detectable with instruments like TPF or Darwin. The simultaneous detection of $O_2$ or $O_3$ and $N_2O$, $CH_4$, or $CH_3Cl$ in the atmosphere of an M-star (or other extrasolar) planet would provide convincing evidence for the existence of extraterrestrial life.

The surfaces of M-star planets also appear to be relatively well shielded from stellar UV radiation, even for the planets circling non-active stars where there is a thinner ozone layer to shield the planetary surface. Thus, the possibility of life on such planets should not be ruled out for this reason.

## ACKNOWLEDGMENTS

This material is based upon work performed by the NASA Astrobiology Institute's Virtual Planetary Laboratory Lead Team, supported by the National Aeronautics and Space Administration through the NASA Astrobiology Institute under Cooperative Agreement Number CAN-00-OSS-01. M.C. thanks NASA for supporting his participation in this work through JPL contract 1234394 and NASA grant NNG04GL49G with the University of California, Berkeley. J.S. acknowledges the support of the NASA Exobiology Program, grant NNG04GK43G. J.F.K. also acknowledges the International Space Science Institute (ISSI) and the support from NASA's Exobiology and Astrobiology Programs. We thank Suzanne Hawley for helpful discussions about M dwarfs and the referees for their valuable suggestions.

## ABBREVIATIONS

HZ, habitable zone; IR, infrared; IUE, International Ultraviolet Explorer; LW, long-wavelength; MIR, mid-infrared; PAL, present atmospheric level; PAR, photosynthetically active radiation; SMART, Spectral Mapping Atmospheric Radiative Transfer; SW, short-wavelength; TPF, Terrestrial Planet Finder; UV, ultraviolet.

Address reprint requests to:
*Antígona Segura*
*California Institute of Technology*
*M/C 220-6*
*Pasadena, CA 91125*

*E-mail:* antigona@ipac.caltech.edu

TABLE 1. STELLAR PARAMETERS

| Star | Effective temperature (K) | Luminosity L/L$_\odot$ | Reference | Planet semi-major axis (AU) |
|---|---|---|---|---|
| AD Leo | 3400 | $2.3 \times 10^{-2}$ | Leggett *et al.* (1996) | 0.16 |
| GJ 643 | 3200 | $4.9 \times 10^{-3}$ | Reid and Gilmore (1984) | 0.07 |
| Non-active star | 3100 | $4.4 \times 10^{-3}$ | Fig. 18, Leggett *et al.* (1996) | 0.07 |

TABLE 2. OZONE COLUMN DEPTH ON PLANETS AROUND M DWARFS

| Parent star | $O_3$ column depth ($cm^{-2}$) |
|---|---|
| Sun | $8.4 \times 10^{18}$ |
| AD Leo | $4.4 \times 10^{18}$ |
| GJ 643 | $1.3 \times 10^{19}$ |
| Quiescent $T_{eff}$ | |
| 3100 K | $1.2 \times 10^{18}$ |
| 3400 K | $2.4 \times 10^{18}$ |
| 3650 K | $3.2 \times 10^{18}$ |

TABLE 3. INCOMING AND SURFACE UV FLUXES FOR PLANETS ORBITING M DWARFS

| | Incoming UV flux (W/m$^2$) | | | | | |
|---|---|---|---|---|---|---|
| | | | Quiescent M dwarfs | | | |
| Spectral range | AD Leo | GJ 643 | $T_{eff}$ = 3100 K | $T_{eff}$ = 3400 K | $T_{eff}$ = 3650 K | Sun |
| 200–400 nm | 0.357 | 0.142 | 0.74 | 2.71 | 4.85 | 105.21 |
| UV-C (<280 nm) | $9.4 \times 10^{-2}$ | $1.9 \times 10^{-2}$ | $4.9 \times 10^{-4}$ | $8.4 \times 10^{-3}$ | $2.6 \times 10^{-2}$ | 69.43 |
| UV-B (280–315 nm) | $3.5 \times 10^{-2}$ | $1.8 \times 10^{-2}$ | $7.4 \times 10^{-3}$ | $5 \times 10^{-2}$ | 0.126 | 15.50 |
| UV-A (315–400 nm) | 0.272 | 0.105 | 0.731 | 2.651 | 4.7 | 82.77 |
| | Surface UV flux (W/m$^2$) | | | | | |
| | | | Quiescent M dwarfs | | | |
| Spectral range | AD Leo | GJ 643 | $T_{eff}$ = 3100 K | $T_{eff}$ = 3400 K | $T_{eff}$ = 3650 K | Sun |
| 200–400 nm | 0.259 | 0.112 | 0.821 | 2.97 | 5.25 | 89.89 |
| UV-C (<280 nm) | $2 \times 10^{-13}$ | $1 \times 10^{-28}$ | $1.2 \times 10^{-7}$ | $1.2 \times 10^{-9}$ | $4.2 \times 10^{-11}$ | $1.2 \times 10^{-21}$ |
| UV-B (280–315 nm) | $5.8 \times 10^{-3}$ | $1.7 \times 10^{-3}$ | $2.7 \times 10^{-3}$ | $1.1 \times 10^{-2}$ | $2.5 \times 10^{-2}$ | 1.33 |
| UV-A (315–400 nm) | 0.253 | 0.110 | 0.818 | 2.96 | 5.22 | 88.55 |

TABLE 4. NORMALIZED SURFACE UV DOSE RATES RELATIVE TO PRESENT EARTH FOR DNA
DAMAGE (VAN BAALEN AND O'DONNEL, 1972) FOR PLANETS AROUND M STARS

| Parent star | UV dose rate |
|---|---|
| Sun | 1.00 |
| AD Leo | $1.44 \times 10^{-2}$ |
| GJ 643 | $7.16 \times 10^{-4}$ |
| Quiescent stars ($T_{eff}$) | |
| 3100 K | $6.05 \times 10^{-2}$ |
| 3400 K | $8.46 \times 10^{-2}$ |
| 3650 K | $9.82 \times 10^{-2}$ |